\documentclass[reprint,superscriptaddress,showpacs,amsmath,amssymb,nofootinbib,aps]{revtex4-2}
\usepackage{graphicx}
\usepackage{dcolumn}
\usepackage{bm}
\usepackage{amsthm,amsmath,amssymb}
\usepackage{mathrsfs}
\usepackage{multirow}
\usepackage[T1]{fontenc}
\usepackage{blindtext}
\usepackage[
colorlinks=true,        
citecolor=blue,         
linkcolor=blue,         
urlcolor=blue           
]{hyperref}             
\usepackage{float}

\begin{document}
\preprint{APS/123-QED}

\title{Exploring Supermassive Compact Dark Matter with the Millilensing Effect of Gamma-Ray Bursts}

\author{Huan Zhou}
\affiliation{Department of Astronomy, School of Physics and Technology, Wuhan University, Wuhan 430072, China}

\author{An Li}
\affiliation{Department of Astronomy, Beijing Normal University, Beijing 100875, China}

\author{Shi-Jie Lin}
\affiliation{Department of Astronomy, Beijing Normal University, Beijing 100875, China}

\author{Zhengxiang Li}
\email{zxli918@bnu.edu.cn}
\affiliation{Department of Astronomy, Beijing Normal University, Beijing 100875, China}
\affiliation{Institute for Frontiers in Astronomy and Astrophysics, Beijing Normal University, Beijing
102206, China}

\author{He Gao}
\email{gaohe@bnu.edu.cn}
\affiliation{Department of Astronomy, Beijing Normal University, Beijing 100875, China}
\affiliation{Institute for Frontiers in Astronomy and Astrophysics, Beijing Normal University, Beijing
102206, China}

\author{Zong-Hong Zhu}
\email{zhuzh@whu.edu.cn}
\affiliation{Department of Astronomy, School of Physics and Technology, Wuhan University, Wuhan 430072, China}
\affiliation{Department of Astronomy, Beijing Normal University, Beijing 100875, China}

\date{\today}

\begin{abstract}
Gravitational lensing effect is one of most significant observational probes to investigate compact dark matter/objects over a wide mass range. In this work, we first propose to derive the population information and the abundance of supermassive compact dark matter in the mass range $\sim10^5-10^7~M_{\odot}$ from 6 millilensed gamma-ray burst (GRB) candidates in 3000 Fermi GRB events using the hierarchical Bayesian inference method. We obtain that, for the mass range $\sim10^5-10^7~M_{\odot}$, the abundance of supermassive compact dark matter is $f_{\rm CO}=10^{-1.60}$ in the log-normal mass distribution scenario. This result is in obvious tension with some other observational constraints, e.g. ultra-faint dwarfs and dynamical friction. However, it also was argued that there is only one system in these 6 candidates has been identified as lensed GRB event with fairly high confidence. In this case, the tension would be significantly alleviated. Therefore, it would be an interesting clue for both the millilensed GRB identification and the formation mechanism of supermassive compact dark matter.
\end{abstract}

\maketitle

\section{Introduction}
The idea that dark matter is in the form of compact objects has a long and controversial history~\cite{Trimble1987}. Theoretical compact dark matter, including the massive compact halo objects (MACHOs), primordial black holes (PBHs), axion mini-clusters, compact mini halos, boson stars, fermion stars and so on, can exist in different mass ranges. In particular, PBH is taken as one of most promissing candidate which could form in the early Universe through the gravitational collapse of primordial density perturbations~\cite{Hawking1971,Carr1974}  (See~\cite{Sasaki2018,Green2021} for recent reviews). Therefore, numerous methods have been proposed to constrain the abundance of compact dark matter, usually referred as the fraction of compact dark matter/objects in dark matter $f_{\rm CO}=\Omega_{\rm CO}/\Omega_{\rm DM}$ at present universe, in various possible mass windows. Moreover, gravitational lensing effects of different kind sources are powerful observational probes to constrain the $f_{\rm CO}$ over a broad mass range from $\mathcal{O}(10^{-10}~M_{\odot})$ to $\mathcal{O}(10^{10}~M_{\odot})$. For example, searching lensed multiple peaks of transients, i.e. fast radio bursts (FRBs), have been proposed to derive constraints on the compact dark matter~\cite{Munoz2016,Laha2020,Liao2021,Zhou2022a,Zhou2022b,Oguri2023,Krochek2022,Connor2023, Kalita2023,Blaes1992,Nemiroff2001,Ji2018,Paynter2021}. 

Gamma ray bursts (GRBs) are extremely energetic explosions occurring in distant galaxies~\cite{Zhang2018book}. Thanks to successful operations of several dedicated detectors, e.g., the Burst And Transient Source Experiment (BATSE) on Compton Gamma Ray Observatory~\cite{Meegan1992}, the Burst Alert Telescope (BAT) on the Neil Gehrels Swift Observatory~\cite{Swift2004}, and the Gamma-Ray Burst Monitor (GBM) on the Fermi Observatory~\cite{Meegan2009}, $\mathcal{O}(10^{4})$ GRBs have been detected and reported. Similar to FRBs, due to prominent observational features including extremely energetic emission at high redshift and high event rate, millilensed GRBs have been proposed as one of most promising probes for searching and constraining compact dark matter for a long time~\cite{Blaes1992, Nemiroff2001,Ji2018,Paynter2021}.

In this paper, on the basis of the well-measured 6 millilensed GRB candidates in 3000 GRBs sample reported by Fermi satellite~\cite{Veres2021,Wang2021,Yang2021,Lin2022}, we first apply the hierarchical Bayesian inference method to investigate the properties of supermassive compact dark matter in the mass range $\sim10^5-10^7~M_{\odot}$. 
This paper is organized as follows. In Section~\ref{sec2}, we introduce the GRB data and the hierarchical Bayesian inference used to derive constraints on the population hyperparameters of supermassive compact dark matter. In Section~\ref{sec3}, we apply this method to the selected GRB observations and present all corresponding results. Finally, conclusions and discussions are presented in Section~\ref{sec4}. Throughout this paper, we use the concordance $\Lambda$CDM cosmology with the best-fitting parameters from the latest $Planck$ cosmic microwave background (CMB) observations~\cite{Planck2018}.

\section{Methodology}\label{sec2}
\subsection{Gamma-Ray Burst Observations}\label{sec2-1}
\begin{figure}
    \centering
     \includegraphics[width=0.48\textwidth, height=0.36\textwidth]{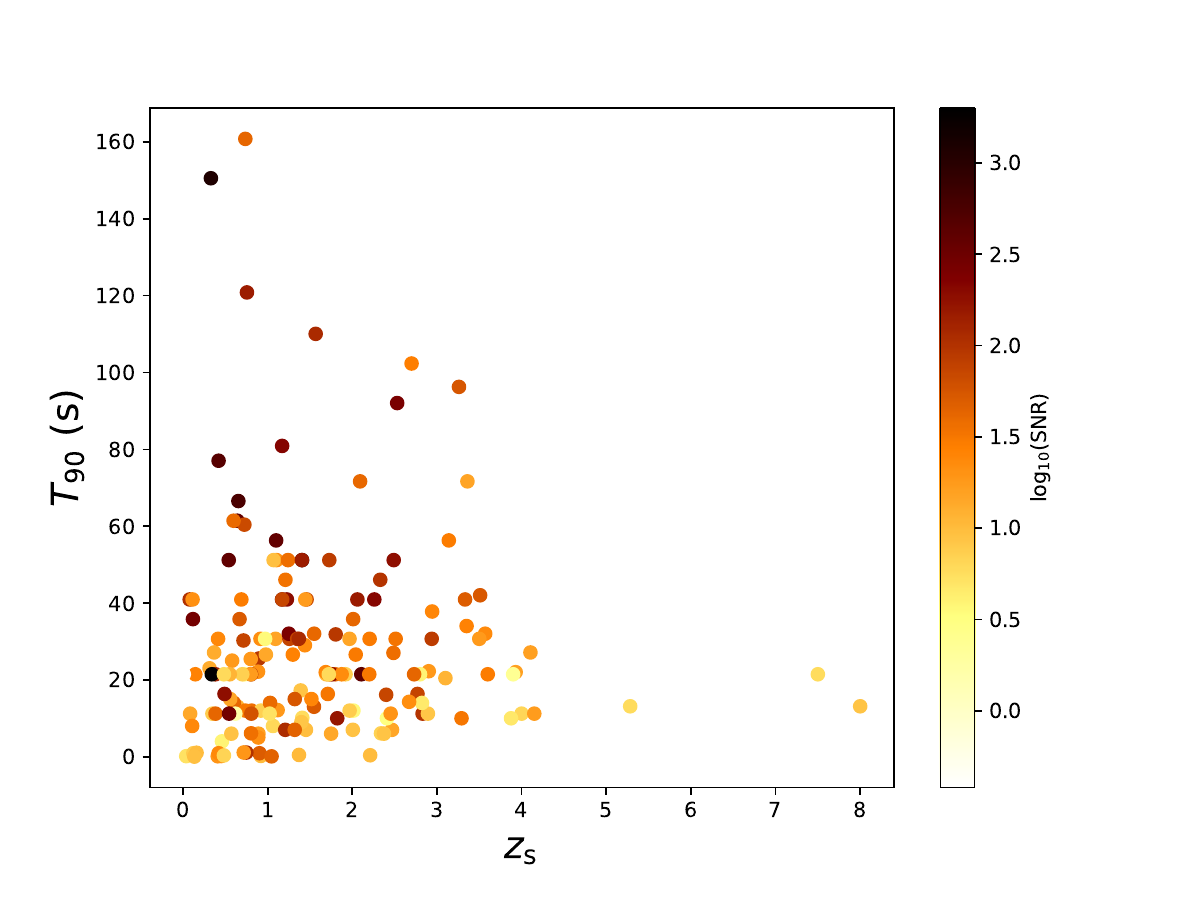}
     \caption{Two-dimensional distribution of redshifts, widths of $T_{90}$ and SNR for the latest 171 GRBs.}\label{fig1}
\end{figure}

In this work, we use the data released by the \textit{Fermi} Gamma Ray Burst Monitor (GBM), which are downloaded from the Fermi Science Support Center's (FSSC) FTP site~\footnote{https://heasarc.gsfc.nasa.gov/FTP/fermi/}, and processed with the GBM data tools v1.1.1. 3000 GRBs were detected by Fermi/GBM up to August 1st, 2022. We use the time-tagged event (TTE) data for our following analysis. The TTE data file records each photon's arrival time with 2 $\mu$s temporal resolution and which of the 128 energy channels the photon registered. In these currently availbale 3000 GRBs, a great deal of effort has been taken to search lensed GRB candidates. Here we collected their results as follow: Yang {\it et al.}~\cite{Yang2021} and Wang {\it et al.}~\cite{Wang2021} independently performed Bayesian analysis of the prompt emission light curves and energy spectra, concluded that GRB 200716C ($M_{\rm L,z}\approx0.43\times10^6~M_{\odot}$) is in favor of the millilensing scenario with two similar pulses. Meanwhile, Veres {\it et al.}~\cite{Veres2021} also carried out exhaustive temporal and spectral analysis for the sample and claimed that GRB 210812A (we use redshifted mass $M_{\rm L,z}\approx1.13\times10^6~M_{\odot}$ from N2 pulse model of GRB210812A, because the evidence using the N2 pulse model is more compelling than in the case of the N1 shape) shows strong evidence in favor of the millilensing effects. Later on, Lin {\it et al.}~\cite{Lin2022} obtained that four candidates pass the hardness test and showed similarities in both temporal and spectral domain, i.e. GRB 081126A ($M_{\rm L,z}\approx5.1\times10^6~M_{\odot}$), GRB 090717A ($M_{\rm L,z}\approx4.02\times10^6~M_{\odot}$), GRB 081122A ($M_{\rm L,z}\approx0.86\times10^6~M_{\odot}$), and GRB 110517B ($M_{\rm L,z}\approx22\times10^6~M_{\odot}$). GRB 081126A and GRB 090717A are ranked as the first-class candidates based on their excellent performance in both temporal and spectrum analysis~\cite{Lin2022}; GRB 081122A and GRB 110517B are ranked as the second-class candidates (suspected candidates), mainly because their two emission episodes show clear deviations in part of the time-resolved spectrum or in the time-integrated spectrum~\cite{Lin2022}. In our following analysis, we firstly assume that all these 6 candidates are millilensed GRB events for exploring properties of compact dark matter. However, it was argued that GRB 090717A, GRB 200716C, GRB 081122A, GRB 081126A, and GRB 110517A can not pass all millilensing tests (light curve similarity test and hardness similarity test)~\cite{Mukherjee2023,Mukherjee2024}. Therefore, we also consider the case that only GRB 210812A in the 6 millilensing candidates is a real lensing system.

For the purpose of this work, we need to collect the distribution of redshift, signal-to-noise ratio (SNR), and the width of each main pulse in the whole GRB sample. In addition, we also need to collect the time delay and the lens mass for all lensing candidates. So firstly we conduct an initial screening of all the events to select GRBs with redshift (usually the redshift is obtained from the spectral lines of afterglow observations) and 171 GRBs are eventually selected. For these events, the time range of the main pulse is selected based on following criteria: 1)SNR of the pulse within the time interval should be greater than 30 and it is calculated as follow: ${\rm SNR}=(C_{\rm all}-C_{\rm bak})/C_{\rm bak}^{1/2}$, in which $C_{\rm all}$ is the photon counts in the time range, $C_{\rm bak}$ is the background photon counts in the time interval; 2)the gap of the pulse should be lower than its 10$\%$ height between other pulses. There are three kinds of situation for the selection. For the first situation, if GRB cannot satisfy the SNR or height criteria, we would choose the whole burst as the main pulse. For the second situation, if there is only one pulse of the GRB satisfying the SNR and height criteria, this pulse would be selected as the main pulse. For the third situation, if there are multiple pulses satisfying the SNR and height criteria, the earliest pulse would be selected as the main one. The width of the main pulse is defined similarly to $T_{90}$, i.e. time interval between $5\%$ and $95\%$ of the cumulative flux. As shown in Figure~\ref{fig1}, we assume that the distribution of width and SNR of the main pulse for whole GRB sample is the same as the 171 GRB sample.

\begin{figure*}
    \centering
    \includegraphics[width=0.48\textwidth, height=0.36\textwidth]{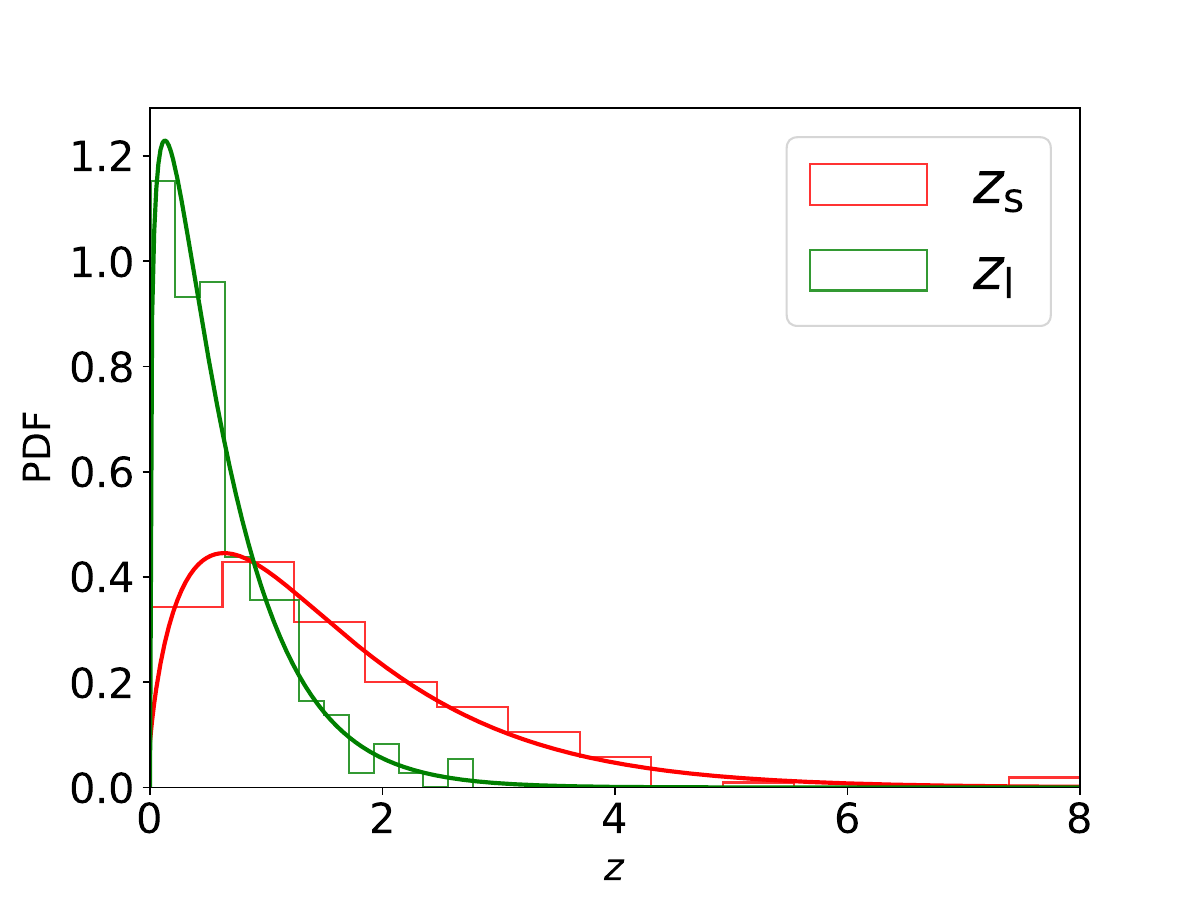}
     \includegraphics[width=0.48\textwidth, height=0.36\textwidth]{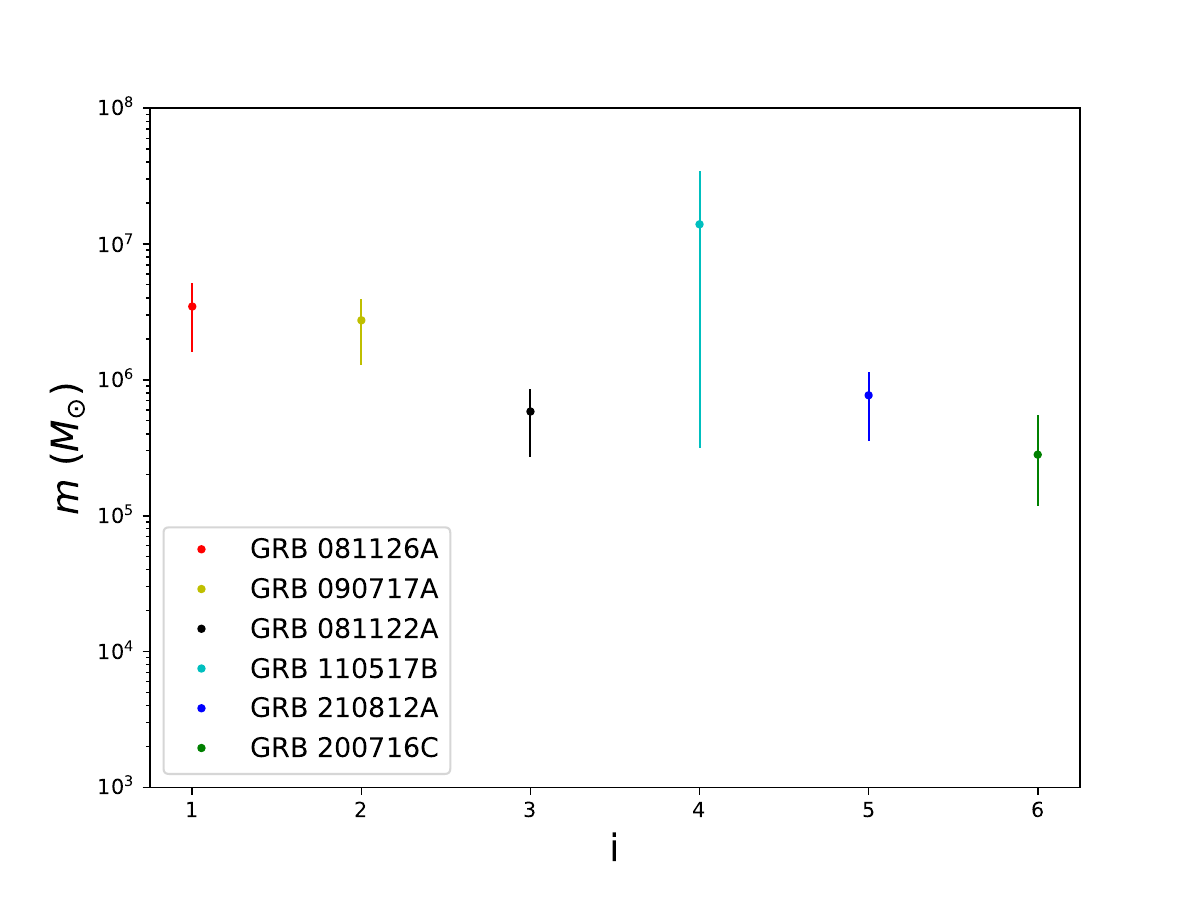}
     \caption{{\bf Left:} Red line represents the redshift distribution of well-measured 171 GRBs. Green line represents the redshift distribution of lens inferred from 171 sources. {\bf Right:} Inferred lens masses of 6 milllesning GRB candidates with $95\%$ confidence levels derived from the posterior distribution of redshifted lens mass $p(M_{z,\rm L}|d_i)$ and redshift distribution of lens.}\label{fig2}
\end{figure*}

If we have the redshift information of these 6 millilensing candidates, we can write the redshift distribution of these lens in the form of the optical depth for each candidate as
\begin{equation}\label{eq4-1}
P_{i}(z_{\rm l})=\frac{1}{\tau(z_{{\rm s},i})}\frac{d\tau(z_{{\rm s},i})}{dz_{\rm l}},
\end{equation}
where $\tau(z_{{\rm s},i})$ is optical depth for each event and is expressed as 
\begin{equation}\label{eq4-1f}
\begin{split}
\tau(z_{{\rm s},i})=\int dm\int_0^{z_{{\rm s},i}}dz_{\rm l} \frac{d n(m,\Phi)}{dm}\times\\
\frac{d\chi(z_{\rm l})}{dz_{\rm l}}(1+z_{\rm l})^2\sigma(m, z_{\rm l},z_{{\rm s},i}),
\end{split}
\end{equation}
where $\chi(z_{\rm l})$, ${d n(m, \Phi)}/{dm}$, and $\sigma(m, z_{\rm l},z_{{\rm s},i})$ are the same as later Eq.~(\ref{eq3-3}) respectively. Unfortunately, we do not have redshift information for these millilensing candidates. As shown in the left panel of Fig.~\ref{fig2}, we can only infer the redshift distribution of lens for 171 GRB sources by combining the redshift distribution of these GRB sources with Eq.~(\ref{eq4-1}). Then we combine this inferred redshift distribution of lens with the posterior distribution $p(M_{z,\rm L}|d_i)$ for each millilensing candidate to infer $p(m|d_i)$ as shown in the right of panel of Fig.~\ref{fig2}. We find that the lens masses of 6 millilensing candidates lie in the range of $10^5-10^7~M_{\odot}$~\footnote{We have tested that the redshift distribution of lenses has little influence on the estimation of magnitude of population information. Therefore, our methods and results should be reliable.}. It is worth noting that here we use the skewnorm distribution to fit the posterior distribution $p(M_{z,\rm L}|d_i)$ obtained from~\cite{Lin2022, Veres2021, Wang2021}.

\begin{table*}[!ht]
\centering
\setlength{\tabcolsep}{13mm}{\begin{tabular}{c|c|c}
\hline
Model & Hyperarameter $\Phi$ & Prior\\
\hline
{} & $\sigma_{\rm c}$  & $\mathcal{U}[0.1,~2]$\\ \cline{2-3}
{Log-normal} & $m_{\rm c}$ & lg-$\mathcal{U}[5,~8]$\\ \cline{2-3}
{} & $f_{\rm CO}$ & lg-$\mathcal{U}[-5,0]$\\
\hline
\end{tabular}}
\caption{\label{tab1} Population hyperarameters $\Phi=[f_{\rm CO}, \sigma_{\rm c}, m_{\rm c}]$ and their prior distributions used in the HBI. $m_{\rm c}$ is in units of $M_{\odot}$.}
\end{table*}

\subsection{Hierarchical Bayesian Inference}\label{sec2-2}
For population hyperparameters of compact dark matter $\Phi=[\boldsymbol p_{\rm mf},f_{\rm CO}]$ and $N_{\rm obs}$ detections of millilensed GRB events $d = [d_1,...d_{N_{\rm obs}}]$, the likelihood follows a Poisson distribution without considering measurement uncertainty and selection effect
\begin{equation}\label{eq3-1}
p(d|\Phi)\propto N(\Phi)^{N_{\rm obs}}e^{-N(\Phi)}.
\end{equation}
However, with measurement uncertainty and selection effect taken into account, the likelihood for $N_{\rm obs}$ milllensed GRB observations can be characterized by the inhomogeneous Poisson process as~\cite{Mandel2019, LVK2022}
\begin{equation}\label{eq3-2}
\begin{split}
&p(d|\Phi)\propto N(\Phi)^{N_{\rm obs}}e^{-N_{\rm det}(\Phi)}\times\\
&\prod_{i}^{N_{\rm obs}}\int d\lambda L(d_i|m)p_{\rm pop}(m|\Phi),
\end{split}
\end{equation}
where the likelihood of one lensing event $L(d_i|m)$ is proportional to the posterior $p(m|d_i)$. $N(\Phi)$ is the total number of millilensing events in the model characterized by the set of population parameters $\Phi$ as
\begin{equation}\label{eq3-3}
\begin{split}
N(\Phi)=\int dm\int dz_{\rm s}\int_0^{z_{\rm s}}dz_{\rm l} \frac{d n(m,\Phi)}{dm}\times\\
\frac{d\chi(z_{\rm l})}{dz_{\rm l}}(1+z_{\rm l})^2\sigma(m, z_{\rm l},z_{\rm s})N_{\rm s}P_{\rm s}(z_{\rm s}),
\end{split}
\end{equation}
where $\chi(z_{\rm l})$ is the comoving distance, ${d n(m, \Phi)}/{dm}$ is the comoving number density of the compact dark matter in a certain extended mass distribution $\psi(m, \boldsymbol p_{\rm mf})$
\begin{equation}\label{eq3-4}
\frac{d n(m, \Phi)}{dm}=\frac{f_{\rm CO}\Omega_{\rm DM}\rho_{\rm c}}{m}\psi(m, \boldsymbol p_{\rm mf}),
\end{equation}
where $\rho_{\rm c}$ is the critical density of the universe. In Eq.~(\ref{eq3-3}), $\sigma(m,z_{\rm l},z_{\rm s})$ is the lensing cross section
\begin{equation}\label{eq3-5}
\sigma(m, z_{\rm l}, z_{\rm s})=\frac{4\pi mD_{\rm l}D_{\rm ls}}{D_{\rm s}}y_0^2.
\end{equation}
It should be emphasized that we take the impact parameter as $y_0=5$ because $y_0>5$ are difficult to be identified as lensing signals\footnote{If the SNR of the secondary peak is greater than 8, the SNR of the main peak is at least greater than 5000. Such a strong GRB signal is almost impossible.}. In addition, $p_{\rm pop}(m|\Phi)$ is the normalized distribution of lens masses and written as
\begin{equation}\label{eq3-6}
p_{\rm pop}(m|\Phi)=\frac{1}{N(\Phi)}\frac{d N(m, \Phi)}{dm}=\psi(m, \boldsymbol p_{\rm mf}).
\end{equation}
Meanwhile, $N_{\rm det}(\Phi)$ is the number of detectable millilensing events and can be defined as 
\begin{equation}\label{eq3-7}
\begin{split}
N_{\rm det}(\Phi)=\int dm\int dz_{\rm s}\int_0^{z_{\rm s}}dz_{\rm l} \frac{d n(m,\Phi)}{dm}\times\\
\frac{d\chi(z_{\rm l})}{dz_{\rm l}}(1+z_{\rm l})^2\sigma_{\rm det}(\lambda, m, z_{\rm l},z_{\rm s})N_{\rm s}P_{\rm s}(z_{\rm s}),
\end{split}
\end{equation}
where $\sigma_{\rm det}(\lambda, m, z_{\rm l},z_{\rm s})$ is the cross section that lensing signal can be detected and depends on the source parameters $\lambda=[{\rm SNR}, w]$ via
\begin{equation}\label{eq3-8}
\begin{split}
\sigma_{\rm det}(\lambda, m, z_{\rm l}, z_{\rm s})=\frac{4\pi mD_{\rm l}D_{\rm ls}}{D_{\rm s}}\times\\
[y_{\max}^2({\rm SNR})-y_{\min}^2(w)].
\end{split}
\end{equation}
The maximum value of the normalized impact parameter $y_{\max}$ can be determined by requiring that the two lensed images are greater than some reference value of flux ratio $R_{\rm f,max}$
\begin{equation}\label{eq3-9}
y_{\rm max}(R_{\rm f,max})=R_{\rm f,max}^{1/4}-R_{\rm f,max}^{-1/4}.
\end{equation}
We take the distribution of SNR in Figure~\ref{fig1} and reference value as function of SNR ($R_{\rm f,max}={\rm SNR}/8$) to obtain $y_{\max}$. In addition, the minimum value of the impact parameter $y_{\min}$ can be determined by the time delay of lensed signals $\Delta t$ and pulse widths $w$
\begin{equation}\label{eq3-10}
\begin{split}
\Delta t(m, z_{\rm l}, y)=4m\big(1+z_{\rm l}\big)\times\\
\bigg[\frac{y}{2}\sqrt{y^2+4}+\ln\bigg(\frac{\sqrt{y^2+4}+y}{\sqrt{y^2+4}-y}\bigg)\bigg]
\geq w.
\end{split}
\end{equation} 
Since the minimum value of impact parameter $y_{\min}$ is not sensitive to the pulse width $w$ as shown in Eq.~(\ref{eq3-10}), we choose the average $\bar{T}_{90}\approx27.6~\rm s$ of 171 GRBs to obtain $y_{\min}$ for our following analysis.
Then the posterior distribution $p(\Phi|d)$ can be calculated from
\begin{equation}\label{eq3-11}
p(\Phi|d)=\frac{p(d|\Phi)p(\Phi)}{Z_{\mathcal{M}}},
\end{equation}
where $p(\Phi)$ is the prior distribution for population hyperparameters $\Phi$ and we assume a log-normal mass function for lenses 
\begin{equation}\label{eq3-12}
\begin{split}
\psi(m, \boldsymbol p_{\rm mf}=[\sigma_{\rm c}, m_{\rm c}])=\frac{1}{\sqrt{2\pi}\sigma_{\rm c} m}\times\\
\exp\bigg(-\frac{\ln^2(m/m_{\rm c})}{2\sigma_{\rm c}^2}\bigg).
\end{split}
\end{equation}
Here, we set prior distributions for all the population hyperparameters $\Phi=[\sigma_{\rm c}, m_{\rm c}, f_{\rm CO}]$ as shown in Tab.~\ref{tab1}. In addition, $Z_{\mathcal{M}}$ is both the normalized factor and the Bayesian evidence for the population model $\mathcal{M}$. This normalized factor can be calculated from the integral of the numerator of Eq.~(\ref{eq3-11}) over $\Phi$, i.e.
\begin{equation}\label{eq3-13}
Z_{\mathcal{M}}=\int d\Phi p(d|\Phi)p(\Phi).
\end{equation}

\section{Results}\label{sec3}
\begin{figure}
    \centering
     \includegraphics[width=0.48\textwidth, height=0.48\textwidth]{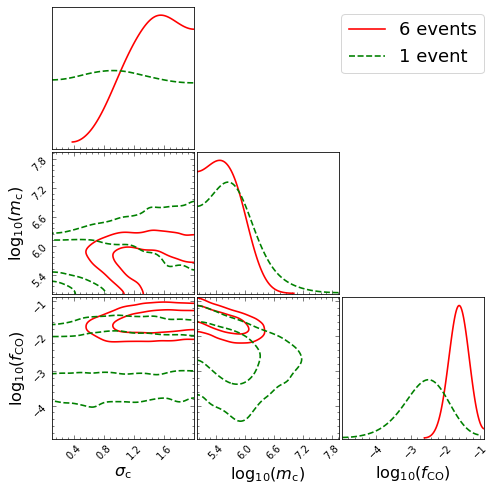}
     \caption{The posterior distributions for hyperparameters $[\sigma_{\rm c}, \log_{10}(m_{\rm c}),\log_{10}(f_{\rm CO})]$ in the log-normal mass function. The red solid line and green dotted line represent the results from 6 millilensing events and 1 millilensing event respectively.}\label{fig3}
\end{figure}
We incorporate the above-mentioned 6 and 1 inferred lens masses from millilensed GRB candidates into the \textbf{EMCEE}~\cite{emcee} with the posterior Eq.~(\ref{eq3-11}) to estimate population hyperparameters $\Phi=[\sigma_{\rm c}, m_{\rm c}, f_{\rm CO}]$ of supermassive compact dark matter, respectively.  Our results are shown in Fig.~\ref{fig3}. For the case including 6 millilensing events, we obtain that the best-fit values and $68\%$ confidence levels for the hyperparameters $[\sigma_{\rm c}, \log_{10}(m_{\rm c}), \log_{10}(f_{\rm CO})]$ are $\sigma_{\rm c}=1.47^{+0.35}_{-0.40}$, $\log_{10}(m_{\rm c})=5.55^{+0.38}_{-0.36}$, $\log_{10}(f_{\rm CO})=-1.60^{+0.23}_{-0.24}$, respectively. For the second case, the best-fit values and $68\%$ confidence levels for the hyperparameters $[\sigma_{\rm c}, \log_{10}(m_{\rm c}), \log_{10}(f_{\rm CO})]$ are $\sigma_{\rm c}=1.03^{+0.64}_{-0.61}$, $\log_{10}(m_{\rm c})=5.71^{+0.43}_{-0.42}$, $\log_{10}(f_{\rm CO})=-2.53^{+0.45}_{-0.62}$, respectively.

\begin{figure}
    \centering
     \includegraphics[width=0.48\textwidth, height=0.36\textwidth]{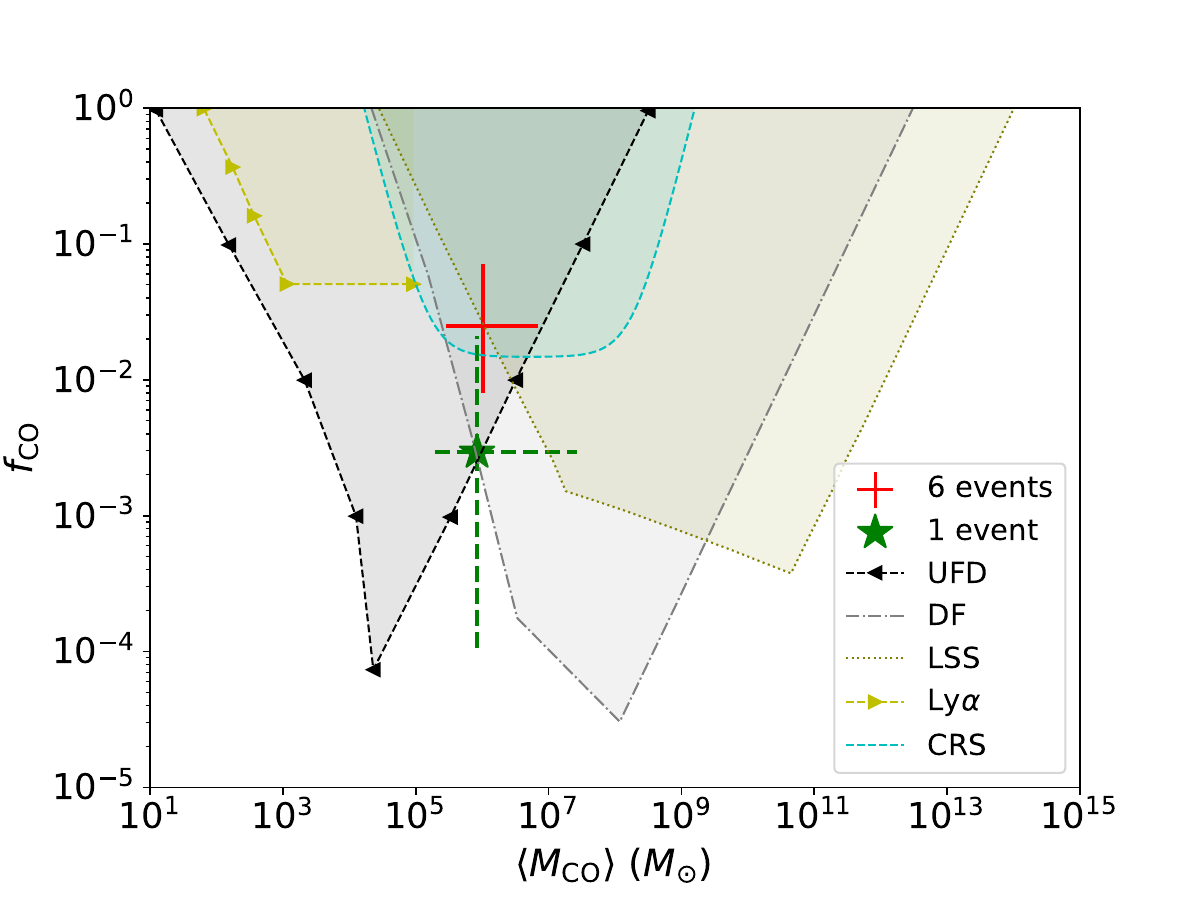}
     \caption{Summary of the constraints on the abundance of compact dark matter $f_{\rm CO}$ with respect to the mean compact dark matter mass $\langle M_{\rm CO}\rangle$. The red cross and green star represent our constraints on $[\langle M_{\rm CO}\rangle, f_{\rm CO}]$ at $95\%$ confidence levels from Fig.~\ref{fig3}. Other lines represent the upper limit of $f_{\rm CO}$ including the ultra-faint dwarfs (UFD)~\cite{Graham2023}, infalling of halo objects due to dynamical friction (DF)~\cite{Carr1999}, cosmic large-scale structure (LSS)~\cite{Carr2018}, Lyman-$\alpha$ forest observations (Ly$\alpha$)~\cite{Murgia2019}, and non-detection of millilensed compact radio sources (CRS)~\cite{Zhou2022c}.}\label{fig4}
\end{figure}

In Fig.~\ref{fig4}, we collect some other currently available and popular constraints on the compact dark matter abundance and compare them to our results from the posterior distributions of the hyperparameters $[\sigma_{\rm c}, \log_{10}(m_{\rm c}), \log_{10}(f_{\rm CO})]$ (red cross), where $\langle M_{\rm CO}\rangle$ is defined as the mean mass of compact dark matter in the log-normal mass function,
\begin{equation}\label{eq4-2}
\langle M_{\rm CO}\rangle=\int m\psi(m, \boldsymbol \sigma_{\rm c}, m_{\rm c})dm=m_{\rm c}e^{\sigma_{\rm c}^2/2}.
\end{equation}
The already existing upper limits are heating the gas of stars through purely gravitational interaction in ultra-faint dwarfs~\cite{Graham2023}, infalling of halo objects due to dynamical friction~\cite{Carr1999}, various cosmic large-scale structure~\cite{Carr2018}, Lyman-$\alpha$ forest observations~\cite{Murgia2019}, non-observation of millilensing compact radio sources (CRSs)~\cite{Zhou2022c}, respectively. There is an obvious tension between our constraint on parameter space $[\langle M_{\rm CO}\rangle, f_{\rm CO}]$ and other upper limits in the mass range $\sim10^5-10^7~M_{\odot}$ for the case of 6 millilensing events as shown in Fig.~\ref{fig4}. However, the case of only one millilensing event significantly alleviates this tension as shown in Fig.~\ref{fig4}.

\section{Conclusions and Discussion}\label{sec4}
In this paper, we have derived constraints on the presence of intergalactic supermassive compact dark matter from 6 miliilensed GRB candidates in 3000 Fermi GRB events. Based on the hierarchical Bayesian inference, we derive the constraint on the fractional abundance of supermassive compact dark matter, $\log_{10}(f_{\rm CO})=-1.60^{+0.23}_{-0.24}$, in the mass range $\sim10^5-10^7~M_{\odot}$. However, there is an obvious tension between our result and some other already existing upper limits at this mass range. Generally, there are several reasons that may lead to this tension, i.e.
\begin{itemize}
\item 1. None of these millilensed GRB candidates has been definitively confirmed as lensing systems so far. If only one of the 6 millilensing candidates is a real lensing system, the abundance of supermassive compact dark matter $f_{\rm CO}$ would reduce to be less than $\sim10^{-2.53}$ and then significantly alleviate this tension in the similar mass range. Moreover, there are some intrinsic burst mechanisms which may cause these similar multi-peak structures instead of lensing effects, for instance, the repeating light-curve properties of these GRBs can be interpreted in the jet precession model~\cite{Gao2023}.

\item 2. If all these events are systems really millilensed by supermassive compact dark matter, in addition to the hypothesis that millilensing candidates of CRSs are confirmed lensing systems~\cite{Casadio2021}, it would be worth considering the special physical mechanisms that produce so many supermassive compact objects, e.g. a scenario that predicts inevitable clustering of PBHs from highly non-Gaussian perturbations has been proposed to produce supermassive PBHs~\cite{Nakama2016,Huang2019,Shinohara2021}, or PBHs growing via accretion~\cite{Ricotti2007} and halo structure~\cite{Delos2023}. Even though in the lensing framework, they may be caused by baryonic matter, such as globular clusters or Population $\rm \uppercase\expandafter{\romannumeral3}$ star.
\end{itemize}
Therefore, whatever these millilensed GRB candidates would be eventually identified, it is foreseen that upcoming complementary multi-messenger observations will yield considerable constraints on both the nature of GRBs and supermassive compact dark matter.

\section{Acknowledgements}
This work was supported by the National Key Research and Development Program of China Grant No. 2021YFC2203001; National Natural Science Foundation of China under Grants Nos.11920101003, 12021003, 11633001, 12322301, and 12275021; the Strategic Priority Research Program of the Chinese Academy of Sciences, Grant Nos. XDB2300000 and the Interdiscipline Research Funds of Beijing Normal University. H.Z is supported by China National Postdoctoral Program for Innovative Talents under Grant No.BX20230271.


\begin{thebibliography}{}
\bibitem{Trimble1987}V. Trimble, Existence and Nature of Dark Matter in the Universe, Ann.Rev.Astron.Astrophys. 25, 425 (1987).

\bibitem{Hawking1971}S. W. Hawking, Gravitationally collapsed objects of very low mass, Mon.Not.Roy.Astron.Soc. 152, 75 (1971).

\bibitem{Carr1974}B. J. Carr, and S.W. Hawking, Black holes in the early Universe, Mon.Not.Roy.Astron.Soc. 168, 399 (1974).

\bibitem{Sasaki2018}M. Sasaki, T. Suyama, T. Tanaka, and S. Yokoyama, Primordial black holes—perspectives in gravitational wave astronomy, Class.Quant.Grav. 35, 063001 (2018).

\bibitem{Green2021}A. M. Green, and B. J. Kavanagh, Primordial Black Holes as a dark matter candidate, J.Phys.G. 48, 043001 (2021).

\bibitem{Munoz2016}J. B. Mu\~{n}oz, E. D. Kovetz, L. Dai, and M. Kamionkowski, Lensing of Fast Radio Bursts as a Probe of Compact Dark Matter, Phys.Rev.Lett. 117, 091301 (2016).

\bibitem{Laha2020}R. Laha, Lensing of fast radio bursts: Future constraints on primordial black hole density with an extended mass function and a new probe of exotic compact fermion and boson stars, Phys.Rev.D. 102, 023016 (2020).

\bibitem{Liao2021}K. Liao, S.-B. Zhang, Z.-X. Li, and H. Gao, Constraints on compact dark matter with fast radio burst observations, Astrophys.J.Lett. 896, L11 (2020).

\bibitem{Zhou2022a}H. Zhou, Z.-X. Li, Z.-Q. Huang, H. Gao, and L. Huang, Constraints on the abundance of primordial black holes with different mass distributions from lensing of fast radio bursts, Mon.Not.Roy.Astron.Soc. 511, 1141 (2022).

\bibitem{Zhou2022b}H. Zhou, Z.-X. Li, K. Liao, C.-H. Niu, H. Gao, Z.-Q. Huang, L. Huang, and B. Zhang, Search for Lensing Signatures from the Latest Fast Radio Burst Observations and Constraints on the Abundance of Primordial Black Holes, Astrophys.J. 928, 124 (2022).

\bibitem{Oguri2023}M. Oguri, V. Takhistov, and K. Kohri, Revealing Dark Matter Dress of Primordial Black Holes by Cosmological Lensing, Phys.Lett.B. 847, 138276 (2023).

\bibitem{Krochek2022}K. Krochek, and E. D. Kovetz, Constraining primordial black hole dark matter with CHIME fast radio bursts, Phys.Rev.D. 105, 103528 (2022).

\bibitem{Connor2023}L. Connor, and V. Ravi, Stellar prospects for FRB gravitational lensing, Mon.Not.Roy.Astron.Soc. 521, 4024 (2023).

\bibitem{Kalita2023}S. Kalita,  S. Bhatporia,  and A. Weltman, Gravitational lensing in modified gravity: a case study for Fast Radio Bursts, JCAP, 11, 059 (2023).

\bibitem{Blaes1992}O. M. Blaes, and R. L. Webster, Using Gamma$\cdot$Ray Bursts to Detect a Cosmological Density of Compact Objects, Astrophys.J.Lett. 391, L63 (1992).

\bibitem{Nemiroff2001}R. J. Nemiroff, G. F. Marani, J. P. Norris, and J. T. Bonnell, Limits on the cosmological abundance of supermassive compact objects from a millilensing search in gamma-ray burst data, Phys.Rev.Lett. 86, 580 (2001).

\bibitem{Ji2018}L.-Y. Ji, E. D. Kovetz, and M. Kamionkowski, Strong Lensing of Gamma Ray Bursts as a Probe of Compact Dark Matter, Phys.Rev.D. 98, 123523 (2018).

\bibitem{Paynter2021}J. Paynter, R. Webster, and E. Thrane, Evidence for an intermediate-mass black hole from a gravitationally lensed gamma-ray burst, Nature Astron. 5, 560 (2021).

\bibitem{Zhang2018book}B. Zhang, The Physics of Gamma-Ray Bursts by Bing Zhang. ISBN: 978-1-139-22653-0. Cambridge Univeristy Press, 2018. doi:10.1017/9781139226530

\bibitem{Meegan1992}C. A. Meegan, G. J. Fishman, R. B. Wilson, et al, Spatial distribution of gamma-ray bursts observed by BATSE, Nature, 355, 143 (1992).

\bibitem{Swift2004}Swift Science Collaboration, The Swift Gamma-Ray Burst Mission, Astrophys.J. 611, 1005 (2004).

\bibitem{Meegan2009}C. Meegan, G. Lichti, P. N. Bhat, et al, The Fermi Gamma-Ray Burst Monitor, Astrophys.J. 702, 791 (2009).

\bibitem{Yang2021}X. Yang, H.-J. Lv, H.-Y. Yuan, et al, Evidence for Gravitational Lensing of GRB 200716C, Astrophys.J.Lett. 921, L29 (2021).

\bibitem{Wang2021}Y. Wang, L.-Y. Jiang, J. Ren, et al, GRB 200716C: Evidence for a Short Burst Being Lensed, Astrophys.J.Lett. 918, L34 (2021).

\bibitem{Veres2021}P. Veres, N. Bhat, N. Fraija, and S. Lesage, Fermi-GBM Observations of GRB 210812A: Signatures of a Million Solar Mass Gravitational Lens, Astrophys.J.Lett. 921, L30 (2021).

\bibitem{Lin2022}S.-J. Lin, A. Li, H. Gao, et al, A Search for Millilensing Gamma-Ray Bursts in the Observations of Fermi GBM, Astrophys.J. 931, 4 (2022).

\bibitem{Planck2018}Planck Collaboration, Planck 2018 results. VI. Cosmological parameters, Astron.Astrophys. 641, A6 (2020).

\bibitem{Mukherjee2023}O. Mukherjee, and R. J. Nemiroff, Light curve and hardness tests for millilensing in GRB 950830, GRB 090717A, and GRB 200716C, Mon.Not.Roy.Astron.Soc. 527, L132 (2024).

\bibitem{Mukherjee2024}O. Mukherjee, and R. Nemiroff, Light curve and hardness tests for millilensing in GRB 081122A, GRB 081126A, GRB 110517B, and GRB 210812A, Mon.Not.Roy.Astron.Soc. 529, L83 (2024).

\bibitem{LVK2022}KAGRA and VIRGO and LIGO Scientific Collaborations, Population of Merging Compact Binaries Inferred Using Gravitational Waves through GWTC-3, Phys.Rev.X. 13, 011048, (2023).

\bibitem{Mandel2019}I. Mandel, W. M. Farr, and J. R. Gair, Extracting distribution parameters from multiple uncertain observations with selection biases, Mon.Not.Roy.Astron.Soc. 486, 1086 (2019). 

\bibitem{emcee}D. Foreman-Mackey, D. W. Hogg,  D. Lang, and J. Goodman, emcee: The MCMC Hammer, Publ.Astron.Soc.Pac. 125, 306 (2013).

\bibitem{Graham2023}P. W. Graham, and H. Ramani, Constraints on Dark Matter from Dynamical Heating of Stars in Ultrafaint Dwarfs. Part 1: MACHOs and Primordial Black Holes, arXiv: 2311.07654.

\bibitem{Carr1999}B. J. Carr, and M. Sakellariadou, Dynamical constraints on dark compact objects, Astrophys.J. 516, 195 (1999).

\bibitem{Carr2018}B. Carr, and J. Silk, Primordial Black Holes as Generators of Cosmic Structures, Mon.Not.Roy.Astron.Soc. 478, 3756 (2018).

\bibitem{Murgia2019}R. Murgia, G. Scelfo, M. Viel, and A. Raccanelli, Lyman-$\alpha$ Forest Constraints on Primordial Black Holes as Dark Matter, Phys.Rev.Lett. 123, 071102 (2019).

\bibitem{Zhou2022c}H. Zhou, Y.-J. Lian, Z.-X. Li, S. Cao, and Z.-Q. Huang, Constraints on the abundance of supermassive primordial black holes from lensing of compact radio sources, Mon.Not.Roy.Astron.Soc. 513, 3627 (2022).

\bibitem{Gao2023}H. Gao, A. Li, W.-H. Lei, and Z.-Q. You, Repeating Emission Episodes in Gamma-Ray Bursts: Millilensing or Jet Precession? Astrophys.J. 945, 17 (2023).

\bibitem{Casadio2021}C. Casadio, D. Blinov, A.C.S. Readhead, et al, SMILE: Search for MIlli-LEnses, Mon.Not.Roy.Astron.Soc. 507, L6 (2021).

\bibitem{Huang2019}Z.-Q. Huang, High-redshift minihaloes from modulated preheating, Phys.Rev.D. 99, 103537 (2019).

\bibitem{Nakama2016}T. Nakama, T. Suyama, and J. Yokoyama, Supermassive black holes formed by direct collapse of inflationary perturbations, Phys.Rev.D. 94, 103522 (2016).

\bibitem{Shinohara2021}T. Shinohara, T. Suyama, and T. Takahashi, Angular correlation as a novel probe of supermassive primordial black holes, Phys.Rev.D. 104, 023526 (2021).

\bibitem{Ricotti2007}M. Ricotti, Bondi accretion in the early universe, Astrophys.J. 662, 61 (2007).

\bibitem{Delos2023}M. S. Delos, and G. Franciolini, Lensing constraints on ultradense dark matter halos, Phys.Rev.D. 107, 083505 (2023).

\end{thebibliography}
\end{document}